# Optimal Medium Access Protocols for Cognitive Radio Networks

Lifeng Lai, Hesham El Gamal, Hai Jiang and H. Vincent Poor

*Abstract*— This paper focuses on the design of medium access control protocols for cognitive radio networks. The scenario in which a single cognitive user wishes to opportunistically exploit the availability of empty frequency bands within parts of the radio spectrum having multiple bands is first considered. In this scenario, the availability probability of each channel is unknown a priori to the cognitive user. Hence efficient medium access strategies must strike a balance between exploring (learning) the availability probability of the channels and exploiting the knowledge of the availability probability identified thus far. For this scenario, an optimal medium access strategy is derived and its underlying recursive structure is illustrated via examples. To avoid the prohibitive computational complexity of this optimal strategy, a low complexity asymptotically optimal strategy is developed. Next, the multi-cognitive user scenario is considered and low complexity medium access protocols, which strike an optimal balance between exploration and exploitation in such competitive environments, are developed.

## I. INTRODUCTION

Recently, the opportunistic spectrum access problem has been the focus of significant research activities [1]. The idea is to allow unlicensed users (i.e., cognitive users) to access the available spectrum when the licensed users (i.e., primary users) are not active, thus to increase the spectral efficiency of the existing wireless networks. The presence of high priority primary users and the requirement that the cognitive users should not interfere with them define a new medium access paradigm that we refer to as *cognitive medium access*. The goal of the current work is to develop a unified framework for the design of efficient, and low complexity, cognitive medium access protocols.

The spectral opportunities available to cognitive users are by their nature time-varying on different time-scales. For example, on a small scale, multimedia data traffic of the primary users will tend to be bursty [2]. On a large scale, one would expect the activities of each user to vary throughout the day. Therefore, to avoid interfering with the primary network, cognitive users must first probe to determine whether there are primary activities before transmission. Under the assumption that each cognitive user cannot access all of the available channels simultaneously [3]–[6], the main task of the medium access protocol is to distributively choose which channels each cognitive user should attempt to use in different time slots, in order to fully (or maximally) utilize the spectral opportunities. The statistical information about the primary users' traffic will be useful for this decision process. For example, with a single cognitive user capable of accessing (sensing) only one channel at each time slot, the problem becomes trivial if the probability that each channel is free is known *a priori*. In this case, the optimal rule is for the cognitive user to access the channel with the highest probability of being free in all time slots. However, such time-varying traffic information is typically not available to the cognitive users *a priori*. The need to learn this information on-line creates a fundamental tradeoff between exploitation and exploration. Exploitation refers to the short-term gain resulting from accessing the channel with the estimated highest probability of being free (based on the results of previous sensing results) whereas exploration is the process by which a cognitive user learns the statistical behavior of the primary traffic (by choosing possibly different channels to probe across time slots). In the presence of multiple cognitive users, the medium access algorithm must also account for the competition between different users over the same channel.

In this paper, we develop a unified framework for the design and analysis of cognitive medium access protocols. This framework allows for the construction of strategies that strike an optimal balance among exploration, exploitation and competition. Tools from reinforcement machine learning are exploited to develop optimal cognitive medium access protocols for the cognitive radio networks. More specifically, we consider the following scenarios in this paper. In the first scenario, we assume the existence of a single cognitive user capable of accessing only a single channel in each time slot. In this setting, we derive an optimal sensing rule that maximizes the expected throughput obtained by the cognitive user. Compared with a genie-aided scheme, in which the cognitive user knows *a priori* the primary network traffic information, there is a throughput loss suffered by any medium access strategy. We obtain a lower bound on this loss and further construct a linear complexity single index protocol that achieves this lower bound asymptotically (when the primary traffic behavior changes very slowly). In the second scenario, we design distributed sensing rules for the scenario in which there are multiple cognitive users. The cognitive users must

L. Lai and H. V. Poor ({llai,poor}@princeton.edu) are with the Department of Electrical Engineering at Princeton University. H. El Gamal (helgamal@ece.osu.edu) is with the Department of Electrical and Computer Engineering at the Ohio State University and is currently visiting Nile University, Cairo, Egypt. H. Jiang (hai.jiang@ece.ualberta.ca) is with the Department of Electrical and Computer Engineering at the University of Alberta. This research was supported by the National Science Foundation under Grants ANI-03-38807 and CNS-06-25637.

also take the competition from other cognitive users into consideration when making sensing decisions. With different assumptions on prior information available at the cognitive users, we develop optimal distributed sensing strategies and characterize the performance loss of these strategies compared with the optimal centralized scheme.

The rest of this paper is organized as follows. Our network model is detailed in Section II. Section III analyzes the scenario in which there is only a single cognitive user. The extension to the multi-user case is reported in Section IV. Finally, Section V summarizes our conclusions and points out several possible future directions. Due to space limitation, we omit the proofs of the results presented in this paper. Interested readers can refer to [7] for details.

## II. NETWORK MODEL

Figure 1 shows the channel model under study. We consider a primary network consisting of $N$ non-overlapping channels, $\mathcal{N} = \{1, \cdots, N\}$, each with bandwidth $B_w$. The users in the primary network are operated in a synchronous time-slotted fashion. We assume that at each time slot, channel $i$ is free with probability $\theta_i$. Let $Z_i(j)$ be a random variable that equals 1 if channel $i$ is free at time slot $j$ and equals 0 otherwise. Hence, given $\theta_i$, $Z_i(j)$ is a Bernoulli random variable with probability density function (pdf)

$$h_{\theta_i}(z_i(j)) = \theta_i \delta(1) + (1 - \theta_i)\delta(0),$$

where $\delta(\cdot)$ is a delta function. Furthermore, for a given $\boldsymbol{\theta} = (\theta_1, \cdots, \theta_N)$, $Z_i(j)$s are independent for each $i$ and $j$. We consider a block varying model in which the value of $\boldsymbol{\theta}$ is fixed for a block of $T$ time slots and then randomly changes at the beginning of the next block according to a joint pdf $f(\boldsymbol{\theta})$.

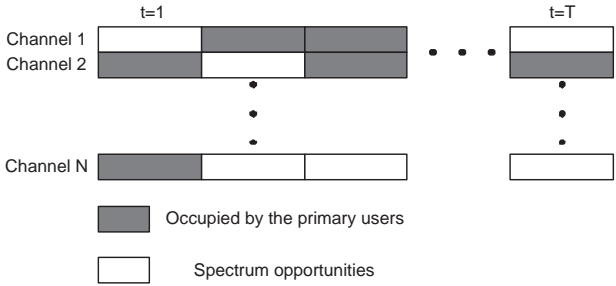

Fig. 1. Channel model.

In our model, the cognitive users attempt to exploit the availability of free channels in the primary network by sensing the activity at the beginning of each time slot. Our work seeks to characterize efficient strategies for choosing which channels to sense (access). The challenge here stems from the fact that the cognitive users are assumed to be unaware of the exact value of $\boldsymbol{\theta}$ *a priori*. We consider two cases in which a cognitive user either has or does not have prior information about the pdf of $\boldsymbol{\theta}$, i.e., $f(\boldsymbol{\theta})$. To further illustrate the point, let us consider our first scenario in which a single cognitive user is capable of sensing only one channel at each time slot. At time slot $j$, the cognitive user selects one channel $S(j) \in \mathcal{N}$ to sense. If the sensing result shows that channel $S(j)$ is free, i.e., $Z_{S(j)}(j) = 1$, the cognitive user can send $B$ bits over this channel; otherwise, the cognitive user will wait until the next time slot and select a possibly different channel to sense. The number of bits that a cognitive user is able to send over a block with $T$ slots is

$$W = \sum_{j=1}^{T} B Z_{S(j)}(j).$$

$W$ is a random variable that depends on the traffic in the primary network and, more importantly for us, on the medium access protocols employed by the cognitive user. Therefore, the overarching goal of Section III is to construct low complexity medium access protocols that maximize $\mathbb{E}\{W\}$. Intuitively, the cognitive user would like to select the channel having the highest probability of being free in order to obtain more transmission opportunities. If $\boldsymbol{\theta}$ is known then this problem is trivial: the cognitive user should choose the channel $i^* = \arg\max_{i \in \mathcal{N}} \theta_i$ to sense. The uncertainty in $\boldsymbol{\theta}$ imposes a fundamental tradeoff between exploration, in order to learn $\boldsymbol{\theta}$, and exploitation, by accessing the channel with the highest estimated availability probability based on current information gathered through sensing, as detailed in the following sections.

## III. SINGLE USER SCENARIO

We start by developing an optimal solution to the single user cognitive user scenario. We can model our single user cognitive medium access problem as a bandit problem, a class of problems studied in reinforcement machine learning. In a typical setting, a decision maker must sequentially choose one process to observe from $N \geq 2$ stochastic processes, which have parameters that are unknown to the decision maker. Associated with each observation is a utility function. The objective of the decision maker is to maximize the sum or discounted sum of the utilities via a strategy that specifies which process to observe for every possible history of selections and observations. A comprehensive treatment covering different variants of bandit problems can be found in [8]–[11].

### A. Optimal Solution for the General Case

The cognitive user employs a medium access strategy $\Gamma$, which will select channel $S(j) \in \mathcal{N}$ to sense at time slot $j$ for any possible causal information pattern obtained through the previous $j - 1$ observations:

$$\Psi(j) = \{s(1), z_{s(1)}(1), \cdots, s(j-1), z_{s(j-1)}(j-1)\}, j \geq 2,$$

i.e. $s(j) = \Gamma(f, \Psi(j))$. Notice that $z_{s(j)}(j)$ is the sensing outcome of the $j$th time slot, in which $s(j)$ is the channel being accessed. If $j = 1$, there is no accumulated information, and thus $\Psi(1) = \phi$ and $s(1) = \Gamma(f)$. We denote the expected value of the payoff obtained by a cognitive user who uses strategy $\Gamma$ as $W_\Gamma = \mathbb{E}_f\{W\}$, where $W$ is defined in Section II.

We further denote
$$V^*(f, T) = \sup_\Gamma W_\Gamma,$$
which is the largest throughput that the cognitive user could obtain when the spectral opportunities are governed by $f(\boldsymbol{\theta})$ and the exact value of each realization of $\boldsymbol{\theta}$ is not known by the user *a priori*. Each medium access decision made by the cognitive user has two effects. The first one is the short term gain, i.e., an immediate transmission opportunity if the chosen channel is found free. The second one is the long term gain, i.e., the updated statistical information about $f(\boldsymbol{\theta})$. This information will help the cognitive user in making better decisions in future stages. There is an interesting tradeoff between the short and long term gains. If we only want to maximize the short term gain, we can pick the one with the highest estimated free probability to sense, based on the current information. This myopic strategy maximally exploits the existing information. On the other hand, by picking other channels to sense, we gain valuable statistical information about $f(\boldsymbol{\theta})$ that can effectively guide future decisions. This process is typically referred to as exploration, as noted previously.

More specifically, let $f^j(\boldsymbol{\theta})$ be the updated pdf after making $j-1$ observations. We begin with $f^1(\boldsymbol{\theta}) = f(\boldsymbol{\theta})$. After observing $z_{s(j)}(j)$, we update the pdf using the following Bayesian formula.

If $z_{s(j)}(j) = 1$,
$$f^{j+1}(\boldsymbol{\theta}) = \frac{\theta_{s(j)} f^j(\boldsymbol{\theta})}{\int \theta_{s(j)} f^j(\boldsymbol{\theta}) d\boldsymbol{\theta}}, \qquad (1)$$

if $z_{s(j)}(j) = 0$,
$$f^{j+1}(\boldsymbol{\theta}) = \frac{\left(1 - \theta_{s(j)}\right) f^j(\boldsymbol{\theta})}{\int \left(1 - \theta_{s(j)}\right) f^j(\boldsymbol{\theta}) d\boldsymbol{\theta}}. \qquad (2)$$

The following result characterizes the optimal medium access control protocols.

*Lemma 1:* For any prior pdf $f$, the following condition specifies $V^*$ and the optimal strategy $\Gamma^*$:
$$V^*(f, T) = \max_{s(1) \in \mathcal{N}} \mathbb{E}_f \left\{ BZ_{s(1)} + V^*\left(f_{Z_{s(1)}}, T-1\right) \right\}, \quad (3)$$
where $f_{Z_{s(1)}}$ is the conditional pdf updated using the Bayesian formula, as if the cognitive user chooses $s(1)$ and observes $Z_{s(1)}$. Also, $V^*\left(f_{Z_{s(1)}}, T-1\right)$ is the value of a bandit problem with prior information $f_{Z_{s(1)}}$ and $T-1$ sequential observations. □

In principle, Lemma 1 provides the solution that maximizes $W_\Gamma$. Effectively, it decouples the calculation at each stage, and hence, allows the use of dynamic programming to solve the problem. The idea is to solve the channel selection problem with a smaller dimension first and then use backward deduction to obtain the optimal solution for a problem with a larger dimension. Starting with $T = 1$, the second term inside the expectation in (3) is 0, since $T - 1 = 0$. Hence, the optimal solution is to choose the channel $i$ having largest $\mathbb{E}_f\{BZ_i\}$, which can be calculated as
$$\mathbb{E}_f\{BZ_i\} = B \int \theta_i f(\boldsymbol{\theta}) d\boldsymbol{\theta}.$$
And $V^*(f, 1) = \max_{i \in \mathcal{N}} \mathbb{E}_f\{BZ_i\}$.

With the solution for $T = 1$ at hand, we can now solve the $T = 2$ case using (3). At first, for every possible choice of $s(1)$ and possible observation $z_{s(1)}$, we calculate the updated pdf $f_{z_{s(1)}}$ using the Bayesian formula. Next, we calculate $V^*(f_{z_{s(1)}}, 1)$ (which is a $T = 1$ problem with updated pdf $f_{z_{s(1)}}$). Finally, applying (3), we have the following equation for the channel selection problem with $T = 2$:
$$V^*(f, 2) = \max_{i \in \mathcal{N}} \int [B\theta_i + \theta_i V^*(f_{z_i=1}, 1) + (1 - \theta_i) V^*(f_{z_i=0}, 1)] f(\boldsymbol{\theta}) d\boldsymbol{\theta}.$$

Correspondingly, the optimal solution is $\Gamma^*(f) = \arg\max_{i \in \mathcal{N}} V^*(f, 2)$, i.e., in the first step, the cognitive user should choose $i^*(1) = \arg\max_{i \in \mathcal{N}} V^*(f, 2)$ to sense. After observing $z_{i^*(1)}$, the cognitive user has $\Psi(1) = \{i^*(1), z_{i^*(1)}\}$, and it should choose $i^*(2) = \arg\max_{i \in \mathcal{N}} V^*(f_{z_{i^*(1)}}, 1)$ implying that $\Gamma^*(f, \Psi(1)) = \arg\max_{i \in \mathcal{N}} V^*(f_{z_{i^*(1)}}, 1)$.

Similarly, after solving the $T = 2$ problem, one can proceed to solve the $T = 3$ case. Using this procedure recursively, we can solve the problem with $T - 1$ observations. Finally, our original problem with $T$ observations is solved as follows.
$$V^*(f, T) = \max_{i \in \mathcal{N}} \int [B\theta_i + \theta_i V^*(f_{z_i=1}, T-1) + (1 - \theta_i) V^*(f_{z_i=0}, T-1)] f(\boldsymbol{\theta}) d\boldsymbol{\theta}.$$

The optimal solution presented above can be simplified when $f(\boldsymbol{\theta})$ has a certain structure, as illustrated by the following examples.

*Example 1:* (One Known Channel) We have $N = 2$ channels with independent primary traffic distributions. Moreover, $\theta_2$ is known. The traffic pattern of channel 1 is unknown, and the probability density function of $\theta_1$ is given by $f_1(\theta_1)$. Since channel 2 is known and is independent of channel 1, sensing channel 2 will not provide the cognitive user with any new information. Hence, once the cognitive user starts accessing channel 2 (meaning that at a certain stage, sensing channel 2 is optimal), there would be no reason to return to channel 1 in the optimal strategy. A generalized version of this assertion was first proved in Lemma 4.1 of [12]. Restating the strategy in our channel selection setup, we have the following lemma.

*Lemma 2:* In the optimal medium access strategy, once the cognitive user starts accessing channel 2, it should keep picking the same channel in the remaining time slots, regardless of the outcome of the sensing process. □

This lemma essentially converts the channel selection problem to an optimal stopping problem [13], where we only need to focus on the strategies that decide at which time-slot we should stop sensing channel 1, if it is ever accessed. The following lemma derives the optimal stopping rule.

*Lemma 3:* For any $f_1(\theta_1)$ and any $T$, if $\theta_2 \geq \Lambda(f_1, T)$, then we should sense channel 2. Here

$$\Lambda(f_1, T) = \max_{\Gamma(f_1)=1} \frac{\mathbb{E}_{f_1}\left\{\sum_{j=1}^{M} Z_1(j)\right\}}{\mathbb{E}_{f_1}\{M\}}, \quad (4)$$

where $\Gamma$ are the set of strategies that start with channel 1 and never switch back to channel 1 after selecting channel 2; and $M$ is a random number that represents the last time slot in which channel 1 is sensed, when the cognitive user follows a strategy in $\Gamma$.

One can now combine Lemma 2 and Lemma 3 to obtain the following optimal strategy.

1) At any time slot $j$, if channel 2 was sensed at time slot $j-1$, keep sensing channel 2.
2) If channel 1 was sensed at time slot $j-1$, update the pdf $f^j$ using (1) and (2) and compute $\Lambda(f_1^j, T-j+1)$ using (4). If $\Lambda(f_1^j, T-j+1) \leq \theta_2$, switch to channel 2; otherwise, keep sensing channel 1. □

*Example 2:* (Independent Channels)

We have $N$ independent channels with $f(\boldsymbol{\theta}) = \prod_{i=1}^{N} f_i(\theta_i)$. This case has a simple form of solution in the asymptotic scenario $T \to \infty$ assuming the following discounted form for the utility function

$$W = \mathbb{E}_f\left\{\sum_{j=1}^{\infty} \alpha^j B Z_{S(j)}(j)\right\},$$

where $0 < \alpha < 1$ is a discount factor. This particular scenario has been considered in [3], and the optimal strategy for this scenario is the following.

1) If channel $l$ was selected at time slot $j-1$, then we get the updated pdf $f_l^j$ using equations (1) and (2), based on the sensing result $z_l(j-1)$. For other channels, we let $f_i^j = f_i^{j-1}, \forall i \neq l, i \in \mathcal{N}$. That is we only update the pdf of the channel which was just accessed (due to the independence assumption).
2) For each channel, we calculate an index using the following equation

$$\Lambda_i(f_i^j) = \max_{\Gamma(f_i^j)=i} \frac{\mathbb{E}_{f_i^j}\left\{\sum_{j=1}^{M} \alpha^j Z_1(j)\right\}}{\mathbb{E}_{f_i^j}\left\{\sum_{j=1}^{M} \alpha^j\right\}},$$

where $\Gamma$ is the set of strategies for the equivalent One-Known-Channel selection problem (with channel $i$ having the unknown parameter) and $M$ is a random number corresponding to the last time slot in which channel $i$ will be selected in the equivalent One-Known-Channel case. $\Lambda_i$ is typically referred to as the Gittins Index [14].
3) Choose the channel with the largest Gittins Index to sense at time slot $j$.

The optimality of this strategy is a direct application of the elegant result of Gittins and Jones [14]. Computational methods for evaluating the Gittins Index $\Lambda$ could be found in [15] and references therein.

### B. Non-parametric Asymptotic Analysis and Asymptotically Optimal Strategies

The optimal solution developed in Lemma 1 suffers from a prohibitive computational complexity. In particular, the dimensionality of our search dimension grows exponentially with the block length $T$. Moreover, one can envision many practical scenarios in which it would be difficult for the cognitive user to obtain the prior information $f(\boldsymbol{\theta})$. In the remaining of this section, we analyze non-parametric schemes that do not explicitly use $f(\boldsymbol{\theta})$, and thus the rules $\Gamma$ considered in the following depend only on $\Psi(j)$ explicitly. We aim to develop schemes that have low complexity but still maintain certain optimality.

For a given strategy $\Gamma$, the expected number of bits the cognitive user is able to transmit through a block with given parameters $\boldsymbol{\theta}$ is

$$\mathbb{E}\{W\} = \sum_{j=1}^{T} B \sum_{i=1}^{N} \theta_i \Pr\{\Gamma(\Psi(j)) = i\}.$$

Recall that $\Gamma(\Psi(j)) = i$ means that, following strategy $\Gamma$, the cognitive user should choose channel $i$ in time slot $j$, based on the available information $\Psi(j)$. Here $\Pr\{\Gamma(\Psi(j)) = i\}$ is the probability that the cognitive user will choose channel $i$ at time slot $j$, following the strategy $\Gamma$.

Compared with the idealistic case where the exact value of $\boldsymbol{\theta}$ is known, in which the optimal strategy for the cognitive user is to always choose the channel with the largest availability probability, the loss incurred by $\Gamma$ is given by

$$L(\boldsymbol{\theta}; \Gamma) = \sum_{j=1}^{T} B\theta_{i^*} - \sum_{j=1}^{T} B \sum_{i=1}^{N} \theta_i \Pr\{\Gamma(\Psi(j)) = i\},$$

where $\theta_{i^*} = \max\{\theta_1, \cdots, \theta_N\}$. We say that a strategy $\Gamma$ is consistent if, for any $\boldsymbol{\theta} \in [0,1]^N$, there exists $\beta < 1$ such that $L(\boldsymbol{\theta}; \Gamma)$ scales as[1] $O(T^\beta)$. The following lemma characterizes the fundamental limits of any consistent scheme.

*Lemma 4:* For any $\boldsymbol{\theta}$ and any consistent strategy $\Gamma$, we have

$$\liminf_{T \to \infty} \frac{L(\boldsymbol{\theta}; \Gamma)}{\ln T} \geq B \sum_{i \in \mathcal{N} \setminus \{i^*\}} \frac{\theta_{i^*} - \theta_i}{D(\theta_i \| \theta_i^*)}, \quad (5)$$

where $D(\theta_i \| \theta_l)$ is the Kullback-Leibler divergence between the two Bernoulli random variables with parameters $\theta_i$ and $\theta_l$ respectively: $D(\theta_i \| \theta_l) = \theta_i \ln(\theta_i/\theta_l) + (1 - \theta_i) \ln((1-\theta_i)/(1-\theta_l))$. □

---

[1] In this paper, we use Knuth's asymptotic notation: 1) $g_1(N) = o(g_2(N))$ means that $\forall c > 0, \exists N_0$, such that $\forall N > N_0, g_1(N) < cg_2(N)$; 2) $g_1(N) = \omega(g_2(N))$ means that $\forall c > 0, \exists N_0$, such that $\forall N > N_0, g_2(N) < cg_1(N)$; 3) $g_1(n) = O(g_2(N))$ means that $\exists c_2 \geq c_1 > 0, N_0$, such that $\forall N > N_0, c_1 g_2(N) \leq g_1(N) \leq c_2 g_2(N)$.

Lemma 4 shows that the loss of any consistent strategy scales at least as $\omega(\ln T)$. An intuitive explanation of this loss is that we need to spend at least $O(\ln T)$ time slots on sampling each of the channels with smaller $\theta_i$, in order to get a reasonably accurate estimate of $\boldsymbol{\theta}$, and hence use it to determine the channel having the largest $\theta_i$ to sense. We say that a strategy $\Gamma$ is order optimal if $L(\boldsymbol{\theta}; \Gamma) \sim O(\ln T)$.

Now, the first question that arises is whether there exist order optimal strategies. As shown later in this section, we can design suboptimal strategies that have loss of order $O(\ln T)$. Thus the answer to this question is affirmative. Before proceeding to the proposed low complexity order-optimal strategy, we first analyze the loss order of some heuristic strategies that may appear to be reasonable in certain applications.

The first simple rule is the random strategy $\Gamma_r$ where, at each time slot, the cognitive user randomly chooses a channel from the available $N$ channels. The fraction of time slots the cognitive user spends on each channel is therefore $1/N$, leading to the loss

$$L(\boldsymbol{\theta}; \Gamma_r) = \frac{B \sum_{i=1}^{N}(\theta_{i^*} - \theta_i)}{N} T \sim O(T).$$

The second one is the myopic rule $\Gamma_g$ in which the cognitive user keeps updating $f^j(\boldsymbol{\theta})$, and chooses the channel with the largest value of

$$\hat{\theta}_i = \int \theta_i f^j(\boldsymbol{\theta}) d\boldsymbol{\theta}$$

at each stage. Since there are no convergence guarantees for the myopic rule, that is $\hat{\boldsymbol{\theta}}$ may never converge to $\boldsymbol{\theta}$ due to the lack of sufficiently many samples for each channel [16], the loss of this myopic strategy is $O(T)$.

The third protocol we consider is *staying with the winner and switching from the loser rule* $\Gamma_{SW}$ where the cognitive user randomly chooses a channel in the first time slot. In the succeeding time-slots 1) if the accessed channel was found to be free, it will choose the same channel to sense; 2) otherwise, it will choose one of the remaining channels based on a certain switching rule.

*Lemma 5:* No matter what the switching rule is, $L(\boldsymbol{\theta}; \Gamma_{SW}) \sim O(T)$. □

Now, we present a linear complexity order optimal strategy.

*Rule 1:* (Order optimal single index strategy) The cognitive user maintains two vectors $\mathbf{X}$ and $\mathbf{Y}$, where each $X_i$ records the number of time slots in which the cognitive user has sensed channel $i$ to be free, and each $Y_i$ records the number of time slots in which the cognitive user has chosen channel $i$ to sense.

1) Initialization: at the beginning of each block, each channel is sensed once.

2) After the initialization period, the cognitive user obtains an estimate $\hat{\boldsymbol{\theta}}$ at the beginning of time slot $j$, given by $\hat{\theta}_i(j) = X_i(j)/Y_i(j)$, and assigns an index

$$\Lambda_i(j) = \hat{\theta}_i(j) + \sqrt{2 \ln j / Y_i(j)}$$

to the $i^{th}$ channel. The cognitive user chooses the channel with the largest value of $\Lambda_i(j)$ to sense at time slot $j$. After each sensing, the cognitive user updates $\mathbf{X}$ and $\mathbf{Y}$. □

*Lemma 6:* The strategy specified in Rule 1 is order optimal. □

The intuition behind this strategy is that as long as $Y_i$ grows as fast as $O(\ln T)$, $\Lambda_i$ converges to the true value of $\theta_i$ in probability, and the cognitive user will choose the channel with the largest $\theta_i$ eventually. The loss of $O(\ln T)$ comes from the time spent on sampling the inferior channels in order to learn the value of $\boldsymbol{\theta}$. This price, however, is inevitable as established in the lower bound of Lemma 4.

## IV. MULTIPLE COGNITIVE USERS SCENARIO

The presence of multiple cognitive users adds an element of competition to the problem. In order for a cognitive user to get hold of a channel now, it must be free from the primary traffic and other competing cognitive users. More rigorously, we assume the presence of a set $\mathcal{K} = \{1, \cdots, K\}$ of cognitive users and consider the distributed medium access decision processes at the multiple users with no coordination. We denote $\mathcal{K}_i(j) \subseteq \mathcal{K}$ as the random set of users who choose to sense channel $i$ at time slot $j$. We assume that the users follow a generalized version of the Carrier Sense Multiple Access/Collision Avoidance (CSMA-CA) protocol to access the channel after sensing the main channel to be free, i.e., if channel $i$ is free, each user $k$ in the set $\mathcal{K}_i(j)$ will generate a random number $t_k(j)$ according to a certain probability density function $g$, and wait the time specified by the generated random number. At the end of the waiting period, user $k$ senses the channel again, and if it is found free, the packet from user $k$ will be transmitted. The probability that user $k$ in the set $\mathcal{K}_i(j)$ gains access to the channel is the same as the probability that $t_k(j)$ is the smallest random number generated by the users in the set $\mathcal{K}_i(j)$. Thus, the throughput user $k$ achieves in a block is

$$W_k = \sum_{j=1}^{T} B Z_{S_k(j)}(j) I \left\{ k = \arg \min_{q \in \mathcal{K}_{S_k(j)}(j)} t_q(j) \right\},$$

in which $S_k(j)$ is the channel selected by the $k^{th}$ user at time slot $j$, and $I(\cdot)$ is an indicator function.

Therefore, user $k$ should devise sensing rule $\Gamma_k$ that maximizes $\mathbb{E}\{W_k\}$. Clearly, even if $\boldsymbol{\theta}$ is known, it is not optimal anymore for all the users to always choose the channel with the largest $\theta_i$ to sense. In particular, if all the users choose the channel with the largest $\theta_i$, the probability that a given user gains control of the channel decreases, while potential opportunities in other channels in the primary network are wasted.

### A. Known $\boldsymbol{\theta}$ Case

To enable a succinct presentation, we first consider the case in which the values of $\boldsymbol{\theta}$ are known to all the cognitive users. The users distributively choose channels to sense and compete for access if the channels are free.

*1) The Optimal Symmetric Strategy:* Without loss of generality, we consider a mixed strategy where user $k$ will choose channel $i$ with probability $p_{k,i}$. Furthermore, we let $\mathbf{p}_k = [p_{k,1}, \cdots, p_{k,N}]$ and consider the symmetric solution in which $\mathbf{p} = \mathbf{p}_1 = \cdots = \mathbf{p}_K$. The symmetry assumption implies that all the users in the network distributively follow the same rule to access the spectral opportunities present in the primary network, in order to maximize the same average throughput each user can obtain. The following result derives the optimal solution in this situation.

*Lemma 7:* For a cognitive network with $K > 1$ cognitive users and $N$ channels with probability $\boldsymbol{\theta}$ of being free, the optimal $\mathbf{p}^*$ is given by

$$p_i^* = \begin{cases} \left\{1 - \left(\frac{\lambda^*}{K\theta_i}\right)^{1/(K-1)}\right\}^+, & \text{for } \theta_i > 0, \\ 0, & \text{for } \theta_i = 0, \end{cases}$$

where $\lambda^*$ is a constant such that $\sum p_i^* = 1$. Here $\{x\}^+ = \max\{0, x\}$. □

The total throughput of the $K$ cognitive users can be represented as

$$\begin{aligned} KW &= BKT \sum \theta_i / K \left\{1 - (1 - p_i^*)^K\right\} \\ &= BT \sum \theta_i \left\{1 - (1 - p_i^*)^K\right\}. \end{aligned}$$

On the other hand, the average total spectral opportunities of the primary network are $BT \sum \theta_i$. This upper bound can be achieved by a centralized channel allocation strategy when $K > N$ (simply by assigning one cognitive user to each channel). Therefore, the loss of the distributed protocol as compared with the centralized scheduling is

$$L = BT \sum \theta_i (1 - p_i^*)^K,$$

If the number of available channels in the network $N$ is fixed and the number of cognitive users $K$ in the network increases, we have the following asymptotic characterization.

*Lemma 8:* Let $2 \leq Q \leq N$ be the number of channels for which $\theta_i > 0$. We have $p_i^* \to 1/Q$, and $L \to 0$ exponentially as $K$ increases, i.e., $L \sim O(e^{-c_1 K})$, where

$$c_1 = \ln(Q/(Q-1)).$$

The reason for the exponential decrease in the loss is that, as the number of cognitive users increases, the probability that there is no user sensing any particular channel decreases exponentially. If $Q = 1$, there is no loss of performance, since all users will always sense the channel with non-zero availability probability.

*2) The Game Theoretic Model:* The optimality of the distributed protocol proposed above hinges on the assumption that all the users will follow the symmetric rule. However, it is straightforward to see that if a single cognitive user deviates from the rule specified in Lemma 7, it will be able to transmit more bits. If this selfish behavior propagates through the network, it may lead to a significant reduction in the overall throughput. This observation motivates our next step in which the channel selection problem is modeled as a non-cooperative game, where the cognitive users are the players, the $\Gamma_k$s are the strategies and the average throughput of each user is the payoff. The following result derives a sufficient condition for the Nash equilibrium in the asymptotic scenario $K \to \infty$.

*Lemma 9:* $(\Gamma_1, \cdots, \Gamma_K)$ is a Nash-equilibrium, if $K$ is large and at each time slot, there are $\tau_i K$ users sensing channel $i$, where $\tau_i$ satisfies $\tau_i = \theta_i / \sum \theta_i$. At this equilibrium, each user has probability $\sum \theta_i / K$ of transmitting at each time slot. □

With this equilibrium result, the cognitive users can use the following stochastic sensing strategy to approximately work on the equilibrium point for a large but finite $K$. Let $s_k(j)$ be the channel chosen by user $k$ at time slot $j$. At each time slot, each user independently selects channel $i$ with probability $\tau_i = \theta_i / \sum \theta_i$, i.e., $\Pr\{s_k(j) = i\} = \tau_i$. Then at each time slot, the number of users sensing channel $i$ will be $\sum_{k=1}^{K} I\{s_k(j) = i\}$, where the $I\{s_k(j) = i\}$s are i.i.d Bernoulli random variables. Hence, the total number of users sensing channel $i$ is a binomial random number, and the fraction of users sensing channel $i$ converges to $\tau_i$ in probability as $K$ increases, i.e.

$$\tau' = \frac{\sum_{k=1}^{K} I\{s_k(j) = i\}}{K} \to \tau_i$$

in probability. Hence, as $K$ increases, the operating point will converge to the Nash equilibrium in probability.

For any $K$, the probability that there is no user choosing channel $i$ to sense is $(1 - \tau_i)^K$. Hence the performance loss compared with the centralized scheme is

$$L = BT \sum \theta_i (1 - \tau_i)^K = BT \sum_{i=1}^{N} \theta_i \left(\frac{\sum_{l=1}^{N} \theta_l - \theta_i}{\sum_{l=1}^{N} \theta_l}\right)^K.$$

It is easy to check that

$$\lim_{K \to \infty} \frac{L}{\exp^{-c_2 K}} = BT\theta_{l^*},$$

where $\theta_{l^*} = \min\{\theta_i : \theta_i > 0\}$, and

$$c_2 = \ln \frac{\sum \theta_i}{\sum_{l=1}^{N} \theta_l - \theta_{l^*}}.$$

It is now clear that the loss of the game theoretic scheme goes to zero exponentially, though the decay rate is smaller than that of the scheme specified in Lemma 7. On the other hand, compared with the scheme in Lemma 7, the game theoretic scheme has the advantage that the cognitive users do not need to know the total number of cognitive users $K$ in the network and, more importantly, they have no incentive to deviate unilaterally.

## B. Unknown $\boldsymbol{\theta}$ Case

Now, we consider the more practical scenario in which $\boldsymbol{\theta}$ is unknown to the cognitive users *a priori*. Hence, the cognitive users also need to estimate $\boldsymbol{\theta}$. Combining the results from single user case and multiple user with known $\boldsymbol{\theta}$ case, we design the following low complexity asymptotically optimal strategy.

*Rule 2:* 1) Initialization: Each user $k$ maintains the following two vectors: $\mathbf{X}_k$, which records the number of time slots in which user $k$ has sensed each channel to be free; and $\mathbf{Y}_k$, which records the number of time slots in which user $k$ has sensed each channel. At the beginning of each block, user $k$ senses each channel once and transmits through this channel if the channel is free and it wins the competition. Also, set $X_{k,i} = 1$, regardless of the sensing result of this stage.

2) At the beginning of time slot $j$, user $k$ estimates $\hat{\theta}_i$ as

$$\hat{\theta}_i(j) = X_{k,i}(j)/Y_{k,i}(j),$$

and chooses each channel $i \in \mathcal{N}$ with probability

$$\hat{\theta}_i(j) / \sum \hat{\theta}_i(j).$$

After each sensing, $\mathbf{X}_k$ and $\mathbf{Y}_k$ are updated. $\square$

*Lemma 10:* If $K$ is large, the scheme in Rule 2 converges to the Nash equilibrium specified in Lemma 9 in probability, as $T$ increases. $\square$

The intuition behind this scheme is that, each user will sample each channel at least $O(T)$ times, and hence as $T$ increases, the estimate $\hat{\boldsymbol{\theta}}$ converges to $\boldsymbol{\theta}$ in probability implying that the unknown $\boldsymbol{\theta}$ case will eventually reduce to the case in which $\boldsymbol{\theta}$ is known to all the users. Hence, if $K$ is sufficiently large, the operating point converges to the Nash equilibrium in probability.

If one can assume that the users will follow the pre-specified rule, then we can design the following strategy that converges to the optimal operating point in probability for any $K$, as $T$ increases.

*Rule 3:* 1) Initialization: Same as Rule 2.

2) At the beginning of time slot $j \leq \ln T$, user $k$ estimates $\hat{\theta}_i$ as

$$\hat{\theta}_i(j) = X_{k,i}(j)/Y_{k,i}(j),$$

and chooses each channel $i \in \mathcal{N}$ with probability

$$\hat{\theta}_i(j) / \sum \hat{\theta}_i(j).$$

For $j \geq \ln T$, the $i^{th}$ channel is sensed with probability

$$\hat{p}_i^* = \left\{ 1 - \left(\lambda^*/\hat{\theta}_i\right)^{1/(K-1)} \right\}^+. \tag{6}$$

After each sensing, $\mathbf{X}_k$ and $\mathbf{Y}_k$ are updated. $\square$

*Lemma 11:* The proposed scheme converges in probability to the optimal operating point specified in Lemma 7, as $T$ increases. $\square$

## V. CONCLUSIONS

This work has developed a unified framework for the design and analysis of cognitive medium access protocols. In the single user scenario, the optimal sensing strategy that balances the tradeoff between exploration and exploitation has been developed. A linear complexity cognitive medium access algorithm, which is asymptotically optimal as the number of time slots increases, has been proposed. The multi-user setting has also been formulated as a competitive bandit problem enabling the design of efficient and game theoretically fair medium access protocols. Our results motivate several interesting directions for future research, for example, developing optimal medium access strategies with consideration of sensing errors and other practical issues. Applying other powerful tools from sequential analysis to design and analyze wireless networks is a promising research direction.